\documentclass{aa}  

\usepackage{graphicx}
\usepackage[varg]{txfonts}
\usepackage{longtable}
\usepackage{txfonts}
\usepackage{bibtex/natbib}
\bibpunct{(}{)}{;}{a}{}{,}

\begin{document}


\newcommand{\adhoc}{{\tt ADHOCw}}
\newcommand{\FP}{Fabry-Perot}
\newcommand{\fantomm}{{\tt FaNTomM}}
\newcommand{\sauron}{{\tt SAURON}}
\newcommand{\XSauron}{{\tt XS}auron}
\newcommand{\tiger}{{\tt TIGER}}
\newcommand{\rotcur}{{\tt ROTCUR}}
\newcommand{\gipsy}{{\tt GIPSY}}
\newcommand{\IRAF}{{\tt IRAF}}
\newcommand{\SN}{$S/N$}
\newcommand{\AN}{$A/N$}
\newcommand{\Ha}{H$\alpha$}
\newcommand{\Hb}{H$\beta$}
\newcommand{\Hg}{H$\gamma$}
\newcommand{\lda}{$\lambda$}
\newcommand{\mgb}{Mg$b$}
\newcommand{\mgt}{Mg$_2$}
\newcommand{\OI}{[{\sc O$\,$i}]}
\newcommand{\OII}{[{\sc O$\,$ii}]}
\newcommand{\OIII}{[{\sc O$\,$iii}]}
\newcommand{\NI}{[{\sc N$\,$i}]}
\newcommand{\NII}{[{\sc N$\,$ii}]}
\newcommand{\NeIII}{[{\sc Ne$\,$iii}]}
\newcommand{\SII}{[{\sc S$\,$ii}]}
\newcommand{\HI}{{\sc H$\,$i}}
\newcommand{\HII}{{\sc H$\,$ii}}
\newcommand{\plm}{$\, \pm \,$}
\newcommand{\Vlos}{$V_\mathrm{los}$}
\newcommand{\Vsys}{$V_\mathrm{sys}$}
\newcommand{\Vrot}{$V_\mathrm{rot}$}
\newcommand{\Vrad}{$V_\mathrm{rad}$}
\newcommand{\epot}{$\varepsilon$}
\newcommand{\czero}{$c_0$}
\newcommand{\sone}{$s_1$}
\newcommand{\cone}{$c_1$}
\newcommand{\stwo}{$s_2$}
\newcommand{\ctwo}{$c_2$}
\newcommand{\sthree}{$s_3$}
\newcommand{\cthree}{$c_3$}
\def\kms{$\mbox{km s}^{-1}$}
\def\kmskpc{$\mbox{km s}^{-1}\mbox{ kpc}^{-1}$}
\def\Myr{$\mbox{M}_\odot\mbox{ yr}^{-1}$}
\def\deg{^\circ}
\def\asim{\mathord{\sim}}
\def\farcs{\hbox{$.\!\!^{\prime\prime}$}}

   \title{Evolution of Structure in Late-type Spiral Galaxies I: Ionized Gas Kinematics in NGC~628}

   \subtitle{}

   \author{K. Fathi\inst{1,2}
          \and 
          J. E. Beckman\inst{1,3}
	  \and 
	  A. Zurita\inst{4}
	  \and
	  M. Rela\~no\inst{4}
	  \and
	  J. H. Knapen\inst{1,5}       
	  \and
	  O. Daigle\inst{6}
	  \and
	  O. Hernandez\inst{6}
	  \and
	  C. Carignan\inst{6}
          }

   \offprints{Kambiz Fathi}

   \institute{
   	      Instituto de Astrof\'\i sica de Canarias, 
   	      C/ V\'\i a L\'actea s/n, 38200 La Laguna, Tenerife, Spain \\
              \email{fathi@iac.es}
	 \and
             Stockholm Observatory, AlbaNova University Center, 
	     106 91 Stockholm, Sweden
	 \and
             Consejo Superior de Investigaciones Cient\'\i ficas (SCIC), Spain
	 \and
             Departamento de F\'{\i}sica Te\'orica y del Cosmos, 
	     Campus de Fuentenueva, Universidad de Granada, E-18071 Granada, Spain
	 \and
             Centre for Astrophysics Research, University of Hertfordshire, 
	     Hatfield, Herts AL10 9AB, UK
	 \and
             Universit\'e de Montr\'eal, C.P. 6128 succ. centre ville,
	     Montr\'eal, QC, Canada H3C 3J7
	     }


   \date{submitted/accepted}

  \abstract
   {}
   {We study two dimensional \FP\ interferometric observations of the nearby face-on late-type spiral galaxy, NGC~628, in order to analyse the ionized gas component of the interstellar medium. Covering the galaxy out to a radius larger than 12 kpc, and with a spatial sampling of 1\farcs6, we aim to investigate the large-scale dynamics as well as feedback from individual \HII\ regions into their surrounding medium.}
   {The observed \Ha\ emission distribution and kinematics are compared with auxiliary data from molecular and atomic gas observations, which display many similarities. We decompose the observed line-of-sight velocities into rotational and higher-order harmonic components, and study the role of gravitational perturbations along with that of external triggers which can disturb the kinematics and morphology of NGC~628. We calculate radial profiles of the emission-line velocity dispersion which we use to study the role of feedback from individual \HII\ regions.}
   {We verify the presence of an inner rapidly rotating disc-like component in NGC~628, which we interpret as caused by slow secular evolution of the large-scale spiral arms and oval structure.  In combination with auxiliary data, we find indication for that gas is falling in from the outer parts towards the central regions, where a nuclear ring has formed at the location of the inner Lindblad resonance radius of an an $m=2$ perturbation. Complementary continuum subtracted narrow band images in \Ha\ have been used to identify 376 \HII\ regions with calibrated luminosities\thanks{The complete version of the \HII\ region catalogue, Table~1 in the online section of this manuscript, is only available in electronic form at the CDS via anonymous ftp to cdsarc.u-strasbg.fr (130.79.128.5) or via http://cdsweb.u-strasbg.fr/cgi-bin/qcat?J/A+A/.}. The mean velocity dispersion for the ionized gas (even when excluding pixels belonging to \HII\ regions) is almost constant out to 12 kpc, although it varies from 14 to 20 \kms, with a steady decline in the outer parts.}
  {We have found kinematic signatures of radial motions caused by an $m=2$ perturbation. Such a perturbation may well be responsible for the inflow of material forming the nuclear ring and the inner rapidly rotating disc-like structure. The latter, in turn, could help build a pseudo-bulge in NGC~628. The current paper demonstrates a number of tools that we have developed for building a solid frame work for studying the evolution of structure in spiral galaxies using two dimensional kinematic observations.}


   \maketitle

\section{Introduction}
Constraining the three dimensional shape of galaxies is one of the most important elements in validating galaxy formation and evolution theories. While images provide projected morphological information, complementary kinematical data is crucial to constrain the shapes and the physical processes that govern them. The vertical components of velocity and velocity dispersion ($\sigma$) are perhaps the most important parameters required to determine the flattening of the various disc layers \citep[e.g.,][]{Bottema2002}, the formation and evolution of bars, bulges, and discs \citep[e.g.,][]{Combes1991, vandenbosch1998}, the intrinsic shape of the dark matter component \citep{zmm1991, bc1997}, and the nature and extent of disc-halo interactions \citep{ni1988}.

If the gaseous layers in the disc are isothermal, both atomic and molecular gas should have similar heights. Accordingly, the atomic and molecular components can be considered as two different phases of the gas coexisting as the same dynamical component, and the vertical oscillations will be of the same amplitude. It is possible that the density of clouds and/or the radiation field at high altitude is not enough to excite the CO molecule, which means that the limit for observing CO will not be coinciding with the limit for neutral hydrogen itself. There is a sharp boundary where the apparent molecular CO fraction falls to zero, while we expect a smoother profile for a unique dynamical gas component. However, due to a lack of kinematic observations of \HII\ regions in galaxies, previous studies of the molecular component have still not been able to verify whether we can observe CO formation and destruction, and therefore put constrains on the velocity and velocity dispersion of the different gaseous components. 

Alternatively, the gas is not isothermal and several layers of gas with different compositions or different thicknesses coexist. The mixing time scales, through atomic or molecular collisions, of two different layers obey the second law of thermodynamics. For a diffuse and homogeneous gas the mixing time scale is of the order of 10$^4$ years. In more realistic cases, the collisional time scale is of the order of 10$^8$ years \citep[e.g.,][]{Cowie1980}. This value is comparable to the dynamical time, but is much larger than the chemical mixing time scale due to the phase change of the gas, estimated to be of the order of 10$^5$ years \citep{lg1989}. Implicitly, the basic constituents of the interstellar medium, i.e., molecular or atomic clouds, cannot be considered as a fluid in equilibrium. If collisions redistribute the kinetic energy, the intrinsic velocity dispersion would be inversely proportional to the cloud mass, whereas in the case of supernova feedback, the resulting gas velocity dispersion is in the range of 4-7 \kms\ throughout the stellar disc \citep{Bottema2003}. Thus, the gas can be stirred up to higher dispersions as was noted by \citet{Gerritsen1997}. Consequently, in the outer parts of the disc, where star formation is less, one might expect a decrease of velocity dispersion \citep{bfq1997}. 

Repeatedly, this picture has been contradicted by observations \citep[e.g.,][]{jimenez-vicenteetal1999}. At large galactocentric radii, where streaming motions should be minimal, the $\sigma$ for giant molecular clouds remains almost constant over several orders of magnitude in brightness. Thus, no equipartition of energy needs to take place and other mechanisms responsible for the heating are required. Relatively small clouds can be heated by star formation and supernovae \citep{cl1981}, whereas the largest clouds could be heated by gravitational scattering \citep{jo1988}. But collective effects such as gravitational instabilities forming structures such as spiral arms, have not yet been taken into account. 

In this context, it is crucial to compare the processes that govern the ionized gas with those that control the molecular and atomic gaseous components. Although the neutral and molecular components are commonly used to derive large-scale kinematic information, higher resolution ionized gas observations are necessary to better study the small-scale velocity structure due to recent star formation  and corresponding feedback processes. Face-on galaxies are favourable targets to derive the local turbulent velocity fields \citep[e.g.,][]{sv1984,cb1997}, since the line width can be attributed almost entirely to the velocity dispersion in the vertical direction. Previous studies have shown a remarkably constant ($\sigma \simeq 7$ \kms) gaseous $\sigma$ as a function of galactocentric radius. This implies that the vertical streaming motions at the arm crossing are not dominant kinematic features. In inclined systems, due to the augmented effect of non-circular motions in the plane of the disc, these studies are complicated \citep[e.g.,][]{vogeletal1993}.

In this paper, we present two-dimensional kinematic maps of the \Ha-emitting gas in the nearly face-on galaxy, NGC~628. We apply a detailed kinematic analysis to explore the role of large-scale versus local ionized gas distribution and dynamics. In Section~\ref{sec:ngc628}, we present a brief review of relevant previous studies of this galaxy, and in Section~\ref{sec:observations} we describe our observations. In Section~\ref{sec:analysis} we present our kinematic analysis tools as well as the results, and finally in Section~\ref{sec:discussion} we discuss our results in the context of secular evolution of discs in late-type spiral galaxies. Finally, in Section~\ref{sec:conclusion}, we conclude the implications of our results.

\section{NGC~628}
\label{sec:ngc628}
NGC~628 (or M74) is an extensively studied isolated grand-design Sc spiral galaxy at a distance of 9.7 Mpc (implying $1\arcsec \simeq 47$ pc), and with a $B$-band magnitude of $M_B$=-19.98 \citep{Tully1988}. The inner disc of NGC~628 is relatively unperturbed, with an inclination of 6.5$^\circ$ and a position angle (PA) of 25$^\circ$ \citep[e.g.,][]{kb1992,cb1997}. This galaxy has a Holmberg radius of 6\arcmin\ with a relatively regular neutral gas velocity field inside this radius \citep[e.g.,][]{wva1986}. From ultraviolet observations, \citet{cornettetal1994} have not been able to confirm the presence of a bulge, but found instead that the nucleus of NGC~628 has the appearance of disc material. \citet{npr1992} had earlier suggested that NGC~628 could be seen as an inner and outer disc, characterised by different stellar populations, with a transition radius located at about 1.5 Holmberg radii. Although the exact location of this transition can be questioned, \citet{cornettetal1994} confirmed this picture, finding that the star formation history of NGC~628 varies with galactocentric distance. Near infrared images show an oval distortion \citep{js1999} and a weak bar \citep{laineetal2002}, of radii 2 kiloparsec (kpc) and 100 pc, respectively. 

Many previous studies have focused on the neutral hydrogen kinematics and distribution \citep[e.g.,][]{sv1984}, the nature, distribution, and composition of \HII\ regions \citep[e.g.,][]{br1992}, and the molecular gas component \citep[e.g.,][]{cb1997}. Neutral hydrogen studies have revealed the presence of an elongated ring-like structure at around 2 Holmberg radii (12\arcmin) distance from the nucleus. This ring lies in a plane with $\simeq$ 15$^\circ$ inclination with respect to the plane of the bright inner disc \citep[e.g.,][]{wa1995}. A strong kinematic PA twist of the \HI\ velocity field indicates that the outer part of the disc, where the ring is observed, is warped. The apparent isolation of NGC~628 makes it difficult to pin down the origin of the warp. The asymmetric gas distribution in the outskirts of the disc, however, could indicate some recent disturbance, as such asymmetric distributions cannot survive for a Hubble time. Alternatively, the detection of two large high-velocity clouds,  symmetrically located on opposite sides of the disc, could explain the warped disc \citep[e.g.,][]{wv1991}. These clouds could be accreting onto the outer parts,  causing the warp \citep{lopezcorredoiraetal2002,beckmanetal2003}. 

This galaxy was also observed with the two-dimensional spectrograph \sauron\  \citep{baconetal2001}. Although the field of view of \sauron\  ($33\arcsec \times 41\arcsec$) is much smaller than those used in all the other studies mentioned above, this instrument delivers stellar as well as ionized gas kinematic information, simultaneously. The stellar $\sigma$ decreases from $\simeq 80$ \kms\ in the outer parts to $\simeq 30$ \kms\ in the central zones, indicating a $<$20\arcsec\ radius dynamically cold central region, maybe an inner disc \citep{gandaetal2006}. Outside this region, the \sauron\ stellar $\sigma$ is similar to that found by  \citet{vf1984}. The latter study shows values of  $60 \pm 20$ \kms\ at about one luminosity scale length, and its evolution with radius is compatible with an exponential decrease, with a radial scale length twice that of the density distribution. This is expected if the stellar disc has a constant scale-height with radius \citep{Bottema1993}. The ionized gas kinematics follows that of the stars. The \Hb\ distribution is more extended than the \OIII\ distribution, both suggesting a ring-like structure of about 15-20\arcsec\ radius \citep{gandaetal2006}. This supports the results from \citet{wa1995}. 

\section{Observations}
\label{sec:observations}
NGC~628 is a member of the large sample of galaxies that were presented in the \FP\ kinematical survey of \citet{daigleetal2006a}. The galaxy was observed with the two dimensional spectrometer \fantomm\ on November 18th 2003, on the 1.6 meter telescope at Observatoire du mont M\'egantic in Qu\'ebec, Canada. \fantomm\ includes a narrow-band (10-20 \AA) interference filter, a \FP\ interferometer and a photon counting imaging device \citep{hernandezetal2003}. The effective focal length over telescope diameter ratio calculated from the effective pixel size on the sky is 2.00. The pixel size is 1\farcs6, with an effective field of view of about $13.7\arcmin\times13.7\arcmin$, thus the data cube delivered by \fantomm\ has the spatial dimensions $512\times512$ pixels.

Our observations of NGC~628 were carried out with a filter centred at $\lambda_c=6598$ \AA, with maximum transmission at $\lambda_c$ and at 293 K of 73\%, and with a Full Width Half Maximum (FWHM) of 18.2 \AA. We scanned 64 channels of 0.11 \AA\ each. This corresponds to a velocity step of 5.2 \kms\ per channel. Each channel exposure was for the duration of 2.33 minutes, adding to a total of 149 minutes effective exposure time. The \FP\ interference order was tuned to $p=899$ at the \Ha\ rest frame wavelength, implying a free spectral range of 333.5 \kms. The \FP\ finesse across the field of view was 23.6, yielding an effective spectral resolution of 21216.

\subsection{Basic Data Reduction}
\label{sec:reduction}
Although the data for this galaxy have been reduced and presented by \citet{daigleetal2006a}, we re-reduced the data in a manner customised for the analysis presented in this paper. Stacking individual exposures of each channel results in three dimensional interferograms. In order to transform these raw interferograms into wavelength-calibrated data cubes, we created two-dimensional phase maps using observations of the narrow Neon emission line at 6598.95 \AA. This line has been chosen as it falls close to the red-shifted \Ha\ line, limiting the phase shift in the dielectric layer of the \FP. We then applied a slight spectral smoothing to the galaxy spectra as well as the calibration spectra by convolving with a Gaussian with a FWHM of 10 \kms, followed by sky subtraction by assuming a uniform OH emission profile throughout the field. Thus, we extracted approximately 20000 spectra, from regions with no \Ha\ emission, evenly spread over the field, averaged all these spectra, and subtracted one representation of the OH emission from the entire data cube. The spectral smoothing is applied in order to improve the derivation of the FWHM, given that, in the low signal regions, non-smoothed data deliver overestimated widths. These parts of the reduction were carried out using the \adhoc\ package \citep{amrametal1998}.

An improved data reduction and preparation package has been presented by \citet{daigleetal2006b}. We tested the robustness of the parameters we have used by cross-checking our results with corresponding files obtained by the latter reduction package, finding no significant differences. In particular, \citet{daigleetal2006b} introduce a two-dimensional adaptive binning method. Although this results in increased \SN\ in the outer parts, we decided not to use the binning since it reduces the spatial resolution in too many regions throughout the observed field. 

\subsection{Deriving the \Ha\ Distribution and Kinematics}
\label{sec:kinder}
We derived the kinematics by fitting one Gaussian profile to each individual emission line. The fits yield the continuum level, the monochromatic \Ha\ intensity, the line-of-sight velocity, and the FWHM for each pixel independently. We estimated the errors for the derived kinematics in a rigorous way by using the bootstrap Monte-Carlo method \citep{pressetal1992}. Accordingly, we fitted the kinematics using Gaussian randomised emission line representations. The errors could be derived in a robust way by repeating this procedure 200 times. In regions with high \SN, we can derive the velocities to within 4 \kms\, and the $\sigma$ to within 5 \kms. These values increase to 7 \kms\ and 8 \kms, respectively, when averaging over all pixels within our field. 

\begin{figure*}
\includegraphics[width=.99\textwidth]{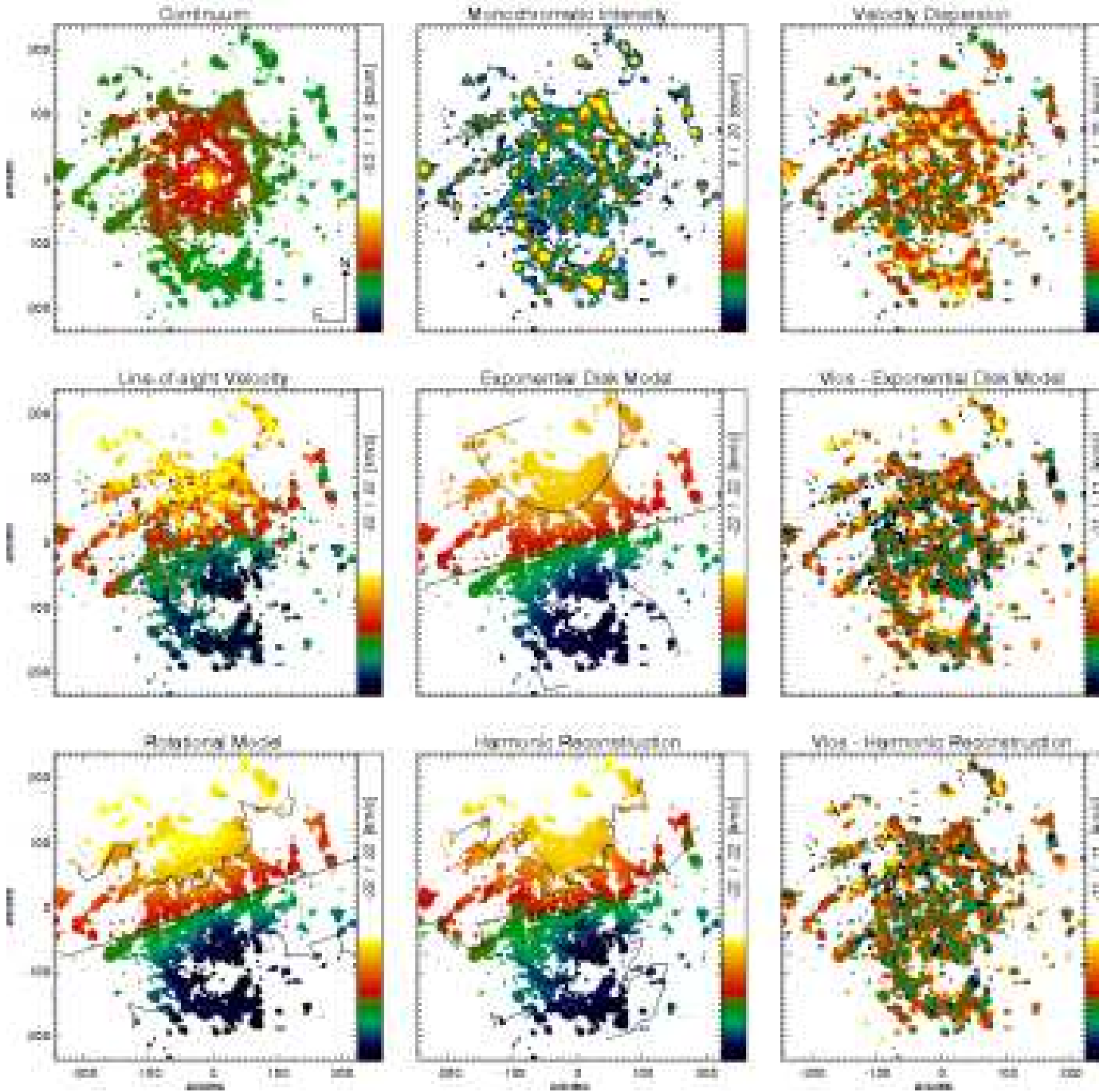}
\caption{The distribution and kinematics of the \Ha\ emission derived from our Gaussian fitting routine as labelled above the individual panels. The north-east direction is indicated at the bottom-right corner of the continuum panel. Also illustrated are the results from our various models for the observed \Ha\ velocity field. All maps have been centred around the kinematic centre (0,0) derived in an iterative way described in Section~\ref{sec:Vanal}. The exponential disc model with corresponding residual field (data - model) is presented in the middle row. The non-parametric purely rotating disc model, the harmonic reconstruction, and the residual (data - harm. rec.) are in the bottom row. The velocity contours are drawn in steps of 20 \kms.}
\label{fig:themaps}
\end{figure*}

We applied a spatial smoothing by convolving each plane of the data cube with a Gaussian to reach a final effective spatial resolution of 3\arcsec. Testing various final resolution values, we found that this values was a good compromise between spatial resolution and coverage, given the intrinsic resolution on the sky of the original data cube. We then calculated the instrumental dispersion using the weighted mean of 10000 calibration spectra, and found $\sigma_\mathrm{instr}$=(FWHM/2.35)=7.37 $\pm$ 0.14 \kms, which we subtract quadratically from the derived emission line $\sigma$ in order to correct for the instrumental broadening. The low error on the derived instrumental dispersion demonstrates that this quantity stays virtually constant across the observed field. To remove any inadequate fits to the spectra, we then employed a number of selection criteria. We removed all pixels for which the amplitude over noise (mean standard deviation of the emission-free region of the spectra) was less than five, or the continuum level was less than zero, or the intrinsic velocity dispersion $\sigma \le  0$. The final result from this cleaning procedure leaves us with 21952 independent kinematic measurements within our field of view (see Fig.~\ref{fig:themaps}).

\subsection{Extracting the Parameters of Individual \HII\ Regions}
\label{sec:HII}

\Ha\ imaging of NGC~628 was obtained with the 1 m Jacobus Kapteyn Telescope on La Palma. The observations were taken on July 6 2003 with photometric sky and  1\farcs2-1\farcs3 seeing. A SITe1 $2048\times2048$ CCD detector was used providing a projected pixel scale of 0\farcs33/pix and a field of view of $\simeq$11\arcmin.

The galaxy was observed through a narrow band (63\AA\ width) interference \Ha\ filter for 2600 s (2 exposures of 900 s plus an exposure of 800 s) and through a Harris $R$-filter for the continuum subtraction for 2$\times$500 s. The width of the \Ha\ filter allows for partial transmission of the nitrogen \NII$\lambda$6548,6583 emission lines. This contribution has not been subtracted from the measured fluxes, and is unlikely to amount to more than 20\% of the total in any specific case.

The data reduction  was carried out using standard \IRAF\footnote{\IRAF\ is distributed by the NOAO, which is operated by AURA under cooperative agreement with the National Science Foundation.} tasks. They include over-scan subtraction and bias and flat field corrections. Afterwards the sky emission was removed from the individual  images by subtracting a constant value obtained from measurements on areas of the image free of galaxy emission. The individual images for the \Ha\ and $R$-band filters were then aligned using the positions of non saturated field stars in both images and combined separately. Then, the on and off-band images of the galaxy were  astrometrically calibrated using the USNO2 catalogue coordinates for the foreground stars of the galaxy images and  the \IRAF\ tasks {\tt ccmap} and {\tt msctpeak}. The accuracy of the astrometric calibration was $\simeq$0.34\arcsec.

Finally the continuum image was subtracted from the on-line (\Ha+continuum) image. To estimate the scaling factor for the continuum subtraction we measured the flux of $\simeq$15 non-saturated foreground stars in both images. The mean value of the flux ratio of on--line  and the off--line filters yielded an initial scale factor of 0.053 with a standard deviation of 0.004. Then we created a set of continuum subtracted images with scale factors ranging from 0.045 to 0.061. A careful inspection of the images showed that the best scaling factor was between 0.049 and 0.053. We finally adopted a factor of 0.051 and estimate that it has an uncertainty of 7\%.

Observations of spectrophotometric standards were not available for our data. Our continuum subtracted image was therefore flux calibrated against previously published \Ha\ photometry for the \HII\ regions of this galaxy \citep{hodge1976,kh1980}. We identified a total of 26 \HII\ regions in both the finding charts of \citet{hodge1976} and our  \Ha\ image. The \HII\ regions were selected to be well distributed across the face of the galaxy and to cover a wide range in fluxes (from $\simeq2\times10^{-14}$ erg\,s\,cm$^{-2}$ to $\simeq3.8\times10^{-13}$ erg\,s\,cm$^{-2}$). This comparison yielded a calibration factor of $(1.22\pm0.05)\times10^{-17}$ erg\,s$^{-1}$\,cm$^{-2}$ for our \Ha\ image, which assuming a distance of 7.3~Mpc \citep{skt1996}, implies a luminosity of $(7.8\pm0.3)\times10^{34}$ erg\,s$^{-1}$ per instrumental count.

\begin{figure*}
\includegraphics[width=.99\textwidth]{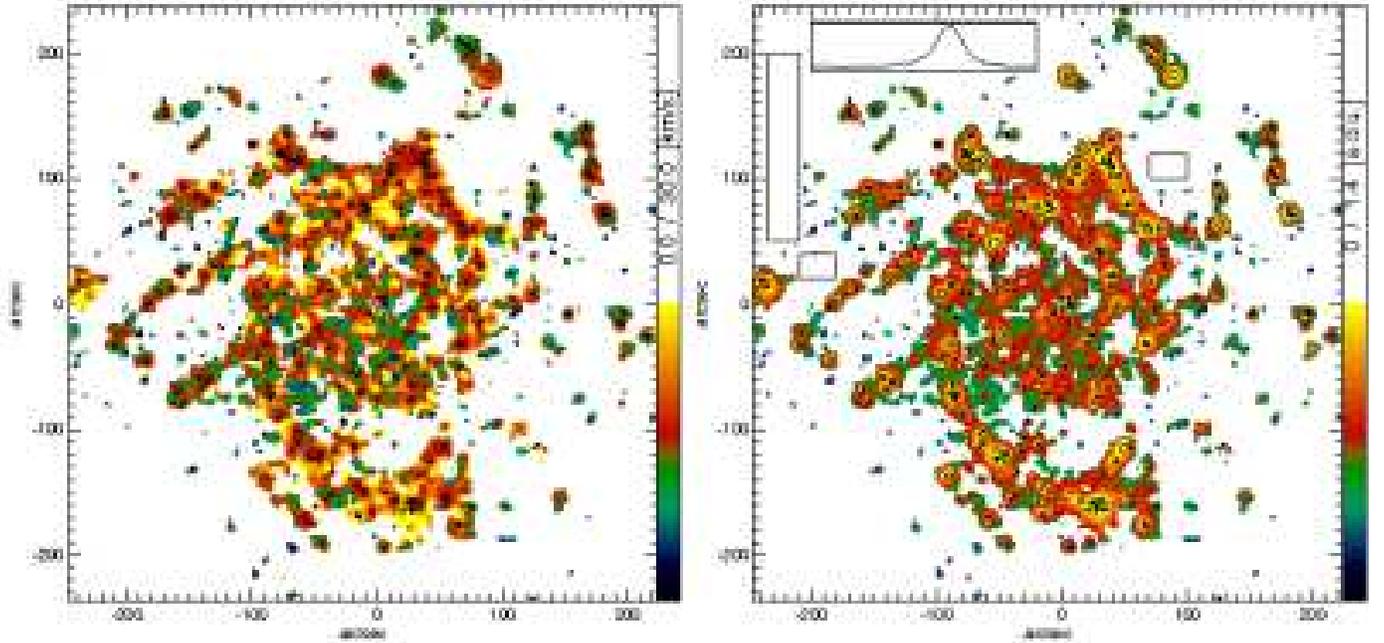}
\caption{The \Ha\ velocity dispersion (left) with over-plotted individual \HII\ regions as described in section~\ref{sec:HII}. The size of each circle indicates the size of the corresponding \HII\ region. The \Ha\ intensity map with circles indicating individual \HII\ regions (right), and rectangles indicating approximate location of regions used to derive the diffuse interstellar gas properties (see section~\ref{sec:discussion}). Spectra for one of these regions, obtained by co-adding the signal from 3731 individual pixels, is illustrated. The centre of the galaxy is marked in both panels.}
\label{fig:sigmamap}
\end{figure*}

The \HII\ region catalogue of NGC~628, i.e., the determination of the position, size and luminosity of the \HII\ regions of the galaxy, was produced with the {\tt REGION} software package developed by C. Heller \citep[see ][ for an application of this code]{zuritaetal2001}. The procedure has been extensively described in a number of papers \citep[e.g.,][]{rozasetal1999,relanoetal2005}. This software has the advantage that it allows for the definition of local background regions above which the \HII\ regions are detected, and for manual editing of the preliminary results, allowing for changes if necessary. The selection criterion for considering a feature in the \Ha\ continuum subtracted image as an \HII\ region is that the feature must contain at least 16 pixels (an area approximately equal to the spatial resolution), each of one having an intensity of at least three times the r.m.s. noise of the image above the local background intensity level. This selection criterion defines the lower luminosity limit for the detection of an \HII\ region in the image, which in this case corresponds to $\log~L_{H\alpha}\sim36.9$~dex. With this selection criteria we detected a total of 376 \HII\ regions in NGC~628, with luminosities ranging from $\log~L_{H\alpha}=36.9$ to 39.3 and diameters up to 316~pc. Our measured fluxes agree within $10-20\%$ with those reported by \citet{kh1980}.  An over plot of the individual \HII\ regions on our observed data is shown in Fig.~\ref{fig:sigmamap}. This catalogue has been made available on the CDS online services\footnote{The \HII\ region catalogue is made available in electronic format at the CDS via anonymous ftp to {\tt cdsarc.u-strasbg.fr} or via {\tt http://cdsweb.u-strasbg.fr/}.}. The CCD \Ha\ image observed at La Palma is not shown in this article.

\section{Kinematic Analysis}
\label{sec:analysis}
\subsection{The Kinematic Analysis Principle}
The first step of our analysis is based on the assumption that the observed gas lies in an infinitesimally thin, and predominantly rotating, circular disc. Observationally, discs show an exponential light distribution, for which the equation for the rotational line-of-sight velocity component was derived by \citet{Freeman1970}. When axisymmetric radial and vertical velocities are present, and taking into account the effect of projection and the convention that positive line-of-sight velocities correspond to recession, the line-of-sight velocities $V_\mathrm{los}$ can be written in polar coordinates as:

\begin{equation}
\label{eq:Vxyz}
\begin{array}{rl}
V_\mathrm{los}(R,\psi, i) &= V_\mathrm{sys}  + V_\mathrm{rot}(R) \cos\psi \sin i \; + \\[4mm]
	&\hspace{2ex} V_\mathrm{rad}(R) \sin\psi \sin i + V_z(R) \cos i,\\
\end{array}
\end{equation}
where \Vsys\ is the systemic velocity of the galaxy, \Vrot\ and \Vrad\ are the rotational and radial velocities, $V_z$ is the vertical velocity component, and $i$ is the inclination at which the disc is projected on the sky. In the case of more inclined systems, where the impact of the radial and rotational velocities on \Vlos\ is dominant, the last term of equation (\ref{eq:Vxyz}) can be ignored \citep[e.g.,][]{zuritaetal2004}. However, in face-on galaxies and in cases where out of plane velocities are large, the effect of $V_z$ can in principle be measured. 

Gas kinematics in real galaxies exhibits radial and/or vertical motions due to, e.g., the presence of bars or spiral arms, which create angle-dependent velocities which cannot be described using equation (\ref{eq:Vxyz}). Hence, additional ingredients in the analysis method are required. \citet{fvdz1994} and \citet{sfdz1997} demonstrated that an observed velocity field can be decomposed into a harmonic series, the coefficients of which provide quantitative information about non-axisymmetric structures: 

\begin{equation}
\label{eq:hardec}
V_\mathrm{los}(R,\phi, i) = c_0 + \sum_{n=1}^{k}{\biggl( c_n \cos (n \phi) + s_n \sin (n \phi)\biggr)\sin i}, \\
\end{equation}
where $c_0$ is the systemic velocity, $c_1$ is the circular velocity, and $s_1$ corresponds to the axisymmetric radial velocity. The angle $\phi$ is measured in the plane of the orbit and is zero at the line of nodes. This formalism has been successfully used by \citet{wbb2004,fathietal2005,emsellemetal2006,krajnovicetal2006,allardetal2006} to model the effects of gravitational perturbations by bars and spiral arms on an axisymmetric velocity field.

\begin{figure*}
\includegraphics[width=.99\textwidth]{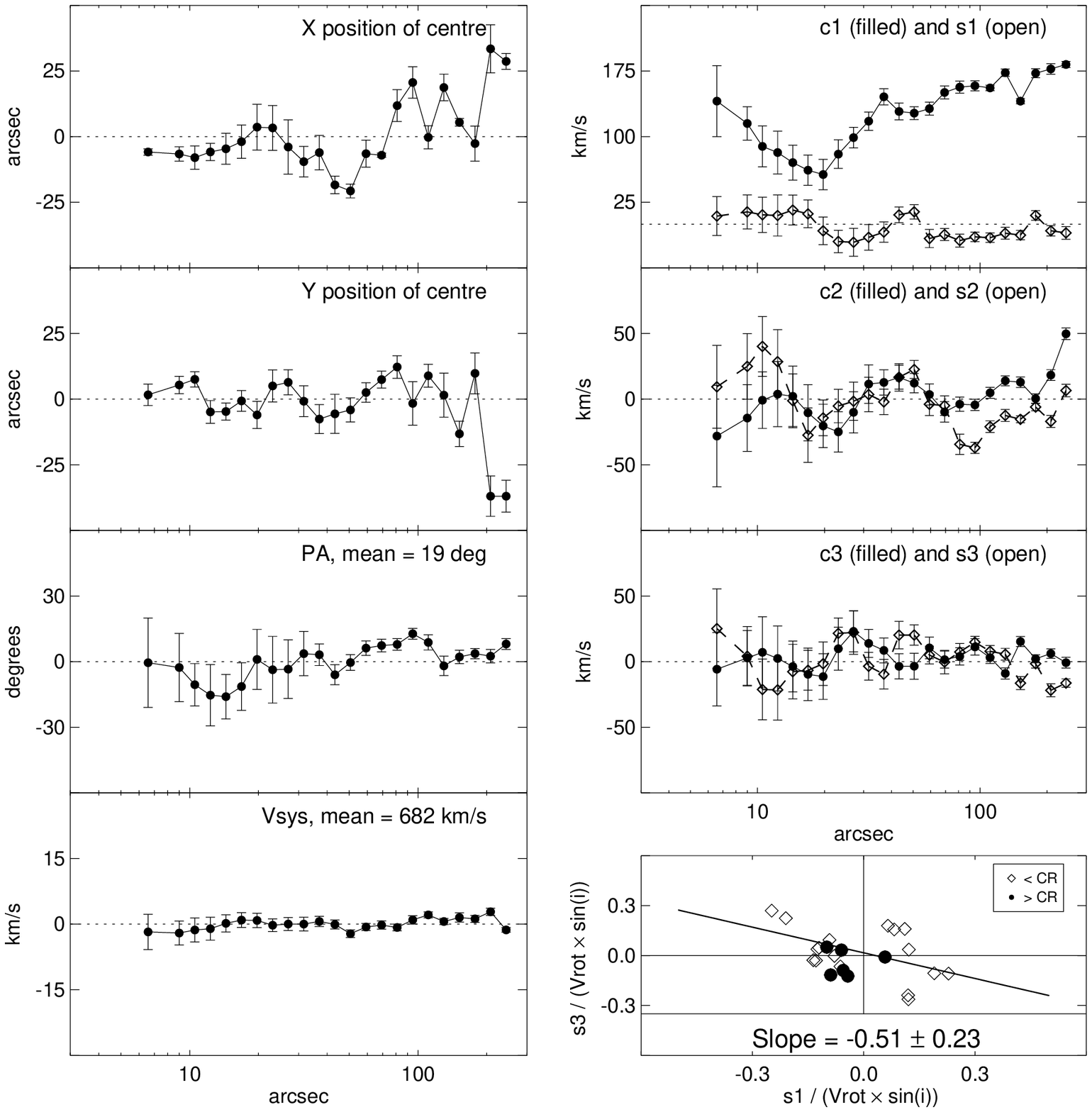}
\caption{The kinematic parameters derived from the velocity field of NGC~628, assuming an inclination of $6.5\deg$. Radial profiles of the kinematic centre, PA, and systemic velocity are presented, the bi-weight mean averages (indicated by the horizontal dashed lines; mean values labelled when appropriate) of which have been calculated to constrain the geometry of the large-scale gaseous disc. The rotation curve and higher harmonic parameters have been derived for the obtained disc geometry, and the error bars were calculated as described in Section~\ref{sec:Vanal}. For a brief discussion of the higher-order parameters ($c_2, s_2, c_3, s_3$), see Section~\ref{sec:analysis}. The bottom right panel shows the $s_3$ versus $s_1$ diagnostics for the region inside the corotation radius of the spiral arms (open symbols), to which we have fitted a line. The slope of the line and its uncertainty is given at the bottom of the panel, where the negative slope indicates the effect of an $m=2$ perturbation inside the corotation radius (see Section~\ref{sec:inflow}).}
\label{fig:parameters}
\end{figure*}

\subsection{Kinematic Analysis Results}
\label{sec:Vanal}

\subsubsection{The Velocity Field}
We started our analysis of the \Ha\ velocity field by fitting an exponential disc model to the observed velocity field. Fixing the centre to the kinematic centre position (see further down in this section), the best-fitting model yields an exponential disc scale length of $64\pm2$\arcsec, kinematic PA$=15\pm 5\deg$, and \Vsys=$683\pm3$ \kms\ (see middle panel of Fig.~\ref{fig:themaps}). Although this fitting method also delivers the central surface density, which can be used to study the mass-to-light ratio of the disc, we find this topic to be beyond the scope of the current paper. We investigated the robustness of this model by omitting regions where we expect large kinematical discrepancies, such as the innermost regions or the outer parts of the arms where the residual velocities are large. We found that the fits delivered very similar parameters every time. Moreover, it is noteworthy that this procedure is computationally very efficient as it delivers the kinematic parameters for all our 21952 observed data points in 1.2 seconds. A computational time much shorter  than the analysis method described next.

An alternative to fitting the entire field simultaneously is to section the field in concentric rings and to derive the kinematic parameters in a non-parametric fashion \citep{Rots1975,bb2001,Teuben2002}. We derived the kinematics according to this method, for a simple rotating model, as well as the \Vlos\ according to the harmonic decomposition in equation (\ref{eq:hardec}). The details of the fitting procedure are presented in \citet{fathietal2005}. The resulting best-fit velocity models are shown in the bottom panels of Fig.~\ref{fig:themaps}. In an attempt to fit the velocity field using equation (\ref{eq:Vxyz}), we found that the radial velocity component is identical to the $s_1$ term of equation (\ref{eq:hardec}), whereas at all radii, $V_z$ stays at zero to within the errors. We conclude that the vertical velocities could not be determined from our data, either because they are too small to be detected with our observations, or because they are mostly driven by small-scale processes, and thus the effects are smeared out within each ring.

It has long been known that erroneous values for the position of the centre, the PA, or the inclination can cause recognisable signatures in the residual velocity fields. Here we will not elaborate on this, but refer the reader to \citet{fathietal2005} for a detailed discussion. In the current study, we have made sure that none of the known systematic features appear in the residual velocity fields presented in Fig.~\ref{fig:themaps}.

Our derivation of the kinematic parameters is done in an iterative way. We first fitted the kinematic centre by calculating the average value of the $(x,y)$ position of the centre. We then calculated the average PA, followed by the average \Vsys, which we derived to be $19\pm7\deg$ and $682\pm2$ \kms, respectively. The averages were calculated using Tukey's bi-weight mean formalism \citep{mt1977} which is particularly advantageous for being insensitive to outliers. Finally, we fitted the harmonic terms of equation (\ref{eq:hardec}). All these parameters and their profiles as a function of galactocentric radius are presented in Fig.~\ref{fig:parameters}. In NGC~628, one clear and one disturbed spiral arm and an oval distortion are the most prominent morphological features, the signatures of which we explore with our kinematical study. As pointed out by \citet{sfdz1997}, the first three terms of equation (\ref{eq:hardec}) should deliver sufficient information for studying gravitational perturbation effects up to and including $m=2$. 

A patchy gas distribution causes a non-uniform distribution of pixels, and it is important to make sure that all the rings include enough pixels to fit harmonic parameters to a reasonable accuracy, especially in the innermost and outermost regions. We adopt a geometric increase of the ring radii with the radial width of the rings increasing by a factor 1 + step, where we adopt a step of $0.15$. This corresponds to an innermost ring radius of 6\farcs5, containing 35 pixels. We only fit the inner 250\arcsec\ of the observed velocity field, since outside this radius the rings are not sufficiently covered and hence do not deliver reliably derived parameters. We found that every ring contained at least 33 data points, which normally leads to reliable results when fitting 6 parameters simultaneously.

We calculate the errors for the tilted-ring and harmonic parameters indicated in Fig.~\ref{fig:parameters}, by means of Monte Carlo simulations. Repeated application of the tilted-ring method to the Gaussian-randomised gas velocity field yields the uncertainties on the harmonic parameters. Our 200 simulations show that choosing geometrically increasing ring radii with steps of $0.15$ indeed yields satisfactory errors. The errors are low, and the parameter profiles are represented by an adequate number of points.

\subsubsection{The Velocity Dispersion}
\begin{figure}
\includegraphics[width=.49\textwidth]{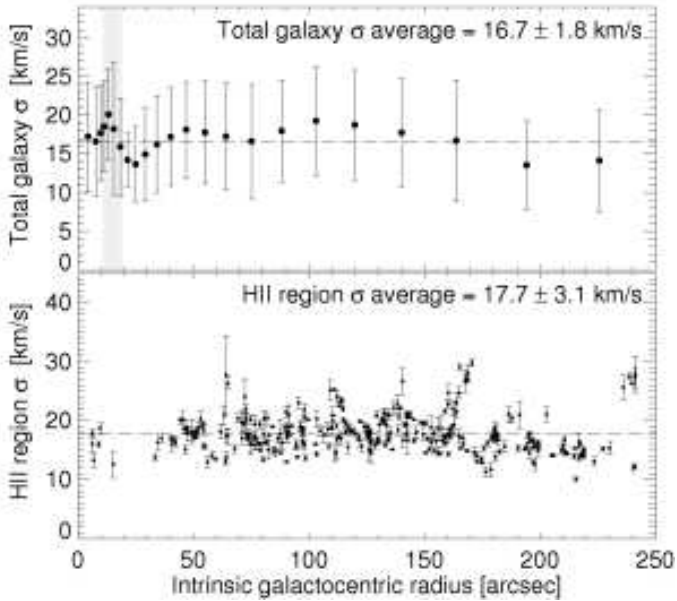}
\caption{The velocity dispersion $\sigma$ corrected for the instrumental dispersion as a function of radius (top panel), and the total $\sigma$ for individual \HII\ regions (bottom panel). The vertical shaded area indicates the position of the nuclear ring-like structure. In each panel, the horizontal dashed line outlines the average value.}
\label{fig:sigmaprofile}
\end{figure}

The large-scale velocity dispersion is one of the key parameters in the study of the physical state of the interstellar medium. For NGC~628, our selection criteria for trusting the Gaussian fits to the spectra allow us to measure reliably the gaseous $\sigma$ intrinsic to the galaxy (see Section~\ref{sec:kinder}). In the same fashion as carrying out the tilted-ring decomposition, we section the $\sigma$ map in order to derive the average of this quantity as a function of galactocentric radius. Fig.~\ref{fig:sigmaprofile} illustrates the average $\sigma$ values in each galactocentric radius bin, where we find that $\sigma$ is almost constant at all radii with an average value of $16.7\pm1.8$ \kms. The average $\sigma$ values reach a peak at around 110\arcsec, followed by a slight decrease outside this radius. The decrease is, however, well within the errors.

The \HII\ region catalogue described in section~\ref{sec:HII} provides positions and sizes for individual \HII\ regions. In order to assess the total contribution of the \HII\ regions to the $\sigma$ profile, we remove all those pixels corresponding to \HII\ regions, and average over radial bins again. We calculate that excluding all \HII\ regions, this quantity remains almost unchanged at $\sigma=16.4\pm1.6$ \kms. The $\sigma$ for the individual \HII\ regions also exhibits no trends as a function of galactocentric radius (see Fig.~\ref{fig:sigmaprofile}). We find that almost all radii within the galaxy include large as well as small \HII\ regions, with an indication of a deficit of \HII\ regions at radii smaller than $\simeq 40$\arcsec, i.e., 1.9 kpc.

\section{Discussion}
\label{sec:discussion}

\subsection{The Global Structure of NGC~628}
We must consider a number of global properties of NGC~628 before building an evolutionary scenario for late-type spiral galaxies of this type. 

It is well established that NGC~628 cannot have undergone any encounter with satellites or other galaxies in the past $10^9$ years \citep{wv1991,kb1992}. The presence of a large-scale elongated structure, an oval, is confirmed from the isophotal analysis by \citet{laineetal2002} at radii around 50\arcsec\ $\simeq 2.3$ kpc. These authors also point out signatures of a nuclear bar at nearly 2\arcsec. This galaxy has one very prominent spiral arm to the south, with one or several disturbed spiral arms to the north, both extending from the nuclear regions out to well beyond the oval. The spiral filaments are possibly disturbed by interaction with the two large high-velocity clouds on either side of the disc \citep[e.g.,][]{kb1992,beckmanetal2003}. \citet{cb1990} found that the star formation efficiency is at its minimum at $\simeq 110\arcsec$, which they interpreted as the corotation radius for the spiral arms. In this galaxy, the O/H ratio has been found to decrease slightly from the centre outwards \citep{Talent1983}, and the \Ha\ luminosity decreases monotonically out to 450\arcsec\ \citep{lr2000}. The oxygen abundance decrease in NGC~628 is rather small, thus moderate mixing of the disc material driven by the oval distortion is allowed \citep{zaritskyetal1994}.

\begin{table}
\caption{The derived deprojected rotation velocities and corresponding errors according to our tilted-ring decomposition.}
\label{tab:rotcurve}
\centering
\begin{tabular}{r r r}
\hline\hline
Radius [arcsec] & \Vrot\ [\kms]& \Vrot\ error [\kms] \\
\hline
   6.5 & 140.5 & 40.5 \\
   8.9 & 115.0 & 19.0 \\
  10.5 &  88.8 & 23.3 \\
  12.3 &  81.7 & 24.4 \\
  14.4 &  70.3 & 19.9 \\
  16.8 &  61.6 & 17.2 \\
  19.7 &  56.6 & 17.3 \\
  23.0 &  79.9 & 16.2 \\
  27.0 &  98.8 & 11.6 \\
  31.5 & 117.6 & 10.4 \\
  36.9 & 145.4 &  9.2 \\
  43.2 & 128.7 &  9.5 \\
  50.6 & 126.8 &  7.3 \\
  59.2 & 131.9 &  7.4 \\
  69.2 & 150.4 &  7.3 \\
  81.0 & 156.5 &  6.9 \\
  94.8 & 158.0 &  5.7 \\
 110.9 & 155.6 &  3.1 \\
 129.8 & 172.9 &  4.2 \\
 151.8 & 140.5 &  3.1 \\
 177.7 & 172.4 &  5.1 \\
 207.9 & 177.3 &  6.0 \\
 243.2 & 182.4 &  3.4 \\
\hline
\end{tabular}
\end{table}

\begin{figure}
\includegraphics[width=.49\textwidth]{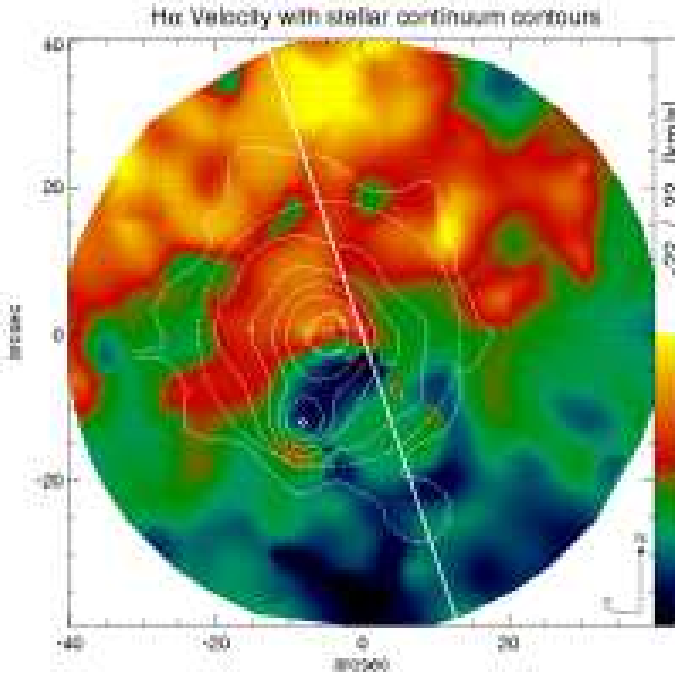}
\caption{Zooming into the central regions of the velocity field of NGC~628. The $HST$ $H$-band image contours (two innermost) and the \fantomm\ continuum image contours (five outermost) are outlined. The kinematic centre and PA=19$\deg$ are marked with a cross and the straight line, respectively. In this figure, linear interpolation has been applied to fill the missing pixels for a clearer illustration. It is clear that the velocities drop at a radius around 20\arcsec, i.e., around the nuclear ring outlined by a dashed ellipse.}
\label{fig:velzoom}
\end{figure}

Gas and stars are often found to be misaligned and/or displaced in spiral galaxies, and the detection of such misalignment has been equally successful in low-inclination and highly inclined systems \citep{Fathi2004,jfbetal2006}. In the case of an external origin of the gas, the nature of the orientation of the gas is thought to depend mostly on the initial angular momentum and mass of the accreted material \citep{qb1992}. In this case, an observed kinematic misalignment can be interpreted as due to accreted gas rotating in a disc which is warped or inclined with respect to the stellar disc \citep{steiman-cameronetal1992}. Analysing high resolution archival {\em Hubble Space Telescope (HST)}, images in the optical and near infra-red, we have found that, within the uncertainties, the photometric PA and centre of NGC~628 are in agreement, to within six degrees and four arcseconds respectively, with the \Ha\ kinematic values for these parameters (see Fig.~\ref{fig:velzoom}).

\begin{figure}
\includegraphics[width=.49\textwidth]{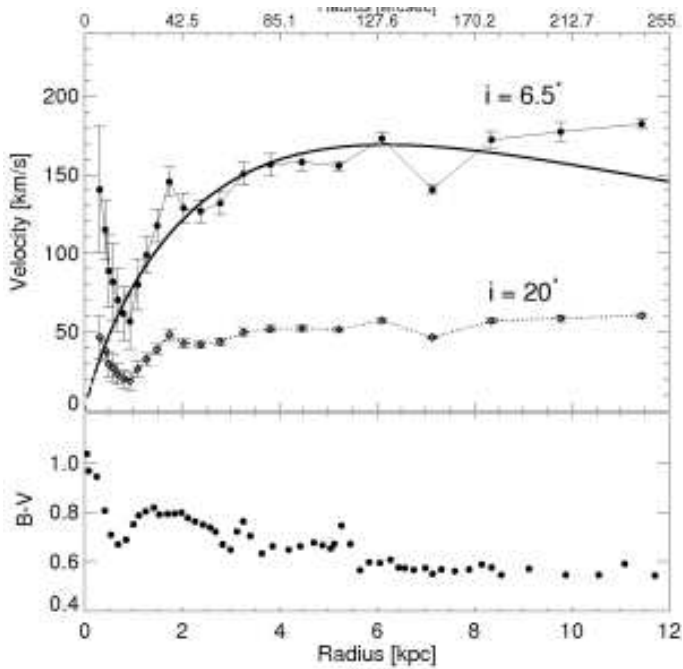}
\caption{Rotation curves assuming two different disc inclinations (top panel) together with an outline of the $B-V$ colour profile from \citet{npr1992} (bottom panel). The solid curve indicates the rotation curve corresponding to a model of an exponential disc profile with a best fit scale length $\sim 3$ kpc, using $6.5\deg$ inclination. See section~\ref{sec:innerparts} for further discussion.}
\label{fig:VrotIncl}
\end{figure}

Figure.~\ref{fig:VrotIncl} shows that the \Ha\ rotation curve rises out to a galactocentric radius of 12 kpc, confirming previous neutral hydrogen observations by \citet{wva1986} (see also \ref{tab:rotcurve}). The curve is quite bumpy, but shows a very clear velocity drop close to around one kpc from the centre. This fall in the rotational velocity coincides quite closely radially with a dip in the $B-V$ index. Although the non-parametric curve (dots with error bars) is not very well fitted by the exponential profile described in Section~\ref{sec:analysis}, we find that an exponential profile could reasonably describe the unperturbed disc component. The exponential mass distribution derived from our velocity field is almost exactly that derived from $R$-band imaging by \citet{npr1992}.  Our velocity field yields a scale length of $\sim64\arcsec$ which is 2.8 kpc on the galaxy, while \citet{npr1992} used their photometric image to derive a scale length of 3.05 kpc, when their distance is converted to ours. 

Comparing only the kinematics of a non-parametric rotation curve with the exponential disc model is not straightforward for the case of NGC~628. In Fig.~\ref{fig:themaps}, we find that the residuals from the exponential disc model do not differ significantly from those yielded by the best fit rotating model. Both residual maps exhibit significant non-circular velocities, in particular in the regions surrounding the prominent spiral arm to the south, and around 50\arcsec\ to the north-east. Although we find that these residuals are taken out by fitting a harmonic decomposition (see bottom row in Fig.~\ref{fig:themaps}), the low inclination of the system, and hence the low amplitudes of the non-circular motions, makes any detailed interpretation difficult. However, following the diagnostics by \citet{wbb2004}, our velocity field confirms that the general overall trend of the third versus the first sinusoidal term follows a negative slope inside the corotation radius (see bottom right panel of Fig.~\ref{fig:parameters}), which suggests that an $m=2$ perturbation is the most prominent in the observed velocity field. This negative slope could be the kinematic confirmation of the presence of the oval perturbation. For this diagnosis, we have used the corotation radius as derived by \citep{cb1990}, and assuming that the pattern speed of the spiral structure is the same as that of the oval structure, based on the response of galaxy disc to an analytic bar or oval potential \citep[e.g., ][]{sh1976} as well as observational and theoretical results for galaxies such as NGC~1068 \citep{emsellemetal2006}, ESO~507-16 \citep{byrdetal1998}, and NGC~1365 \citep{lla1996}. 

The $\sigma$ map presented in Fig.~\ref{fig:sigmamap} displays local variations throughout the observed field, in comparison with the location of the individual \HII\ regions. We calculate the $\sigma$ of individual \HII\ regions by quadratically adding all the spectra within each \HII\ region, to which we then fit a single Gaussian profile to derive the kinematics. Analysing the dependence of the $\sigma$ with galactocentric radius, shows that this quantity stays almost unchanged all the way out to 250\arcsec, regardless of whether we take into account the individual \HII\ regions or not (see Fig.~\ref{fig:sigmaprofile}). This is not unexpected given that less than 18\% of all pixels belong to individual \HII\ regions, and their average $\sigma$ is not considerably higher than the mean $\sigma$ throughout the disc, which is 16.7 \kms.

The constancy with radius of the ionized \Ha\ emitting gas velocity dispersion is analogous to the constancy of $\sigma$ for CO \citep{cb1997} and also for \HI\ \citep{kb1992}. While the densities of the gaseous components are decreasing with radius, their dispersions vary very little. This is an interesting kinematic constraint on overall structure. The fact that the $\sigma$ for the diffuse ionized component is larger than those of CO and \HI\ is consistent with a larger scale height for this component, and this is as might be expected from detailed work in the Milky Way by Reynolds and his group \citep[e.g.,][]{mr2005}. We will discuss this further in section~\ref{sec:difflayers}.

We derived the $\sigma$ of the diffuse interstellar by carefully selecting regions within the field with no significant \Ha\ emission line, i.e., with emission line amplitude-over-noise less than 3 (see Fig.~\ref{fig:sigmamap}). The spectra from sections of these regions were then co-added quadratically, and a single Gaussian was fitted to the result. We confirm that the velocities for these selected region are in agreement with Fig.~\ref{fig:themaps}. The emission line $\sigma$, on the other hand, was found to be similar to that of the ionized gas. For the diffuse gas, we derive the weighted average velocity dispersion $\sigma=16.3\pm1.3$ \kms. This value is significantly lower than that found by many studies such as \citet{whl1997} who derived for five late-type spirals, the diffuse \Ha\ average velocity dispersion larger than $\simeq 30$ \kms. The value we have derived shows that our data cannot be used to properly probe the diffuse interstellar medium.  

\subsection{The innermost kiloparsec}
\label{sec:innerparts}
In Fig.~\ref{fig:VrotIncl}, we see that the rotation curve of NGC~628 starts at high values, drops to a minimum  just before a radius of 20\arcsec, corresponding to 0.9 kpc (n.b. $1\arcsec \simeq 47$ pc), then rises again. Under normal conditions and mass-to-light ratios, such a central rise and fall is indicative of a rapidly rotating compact central structure, in this case smaller than 20\arcsec, which fits well with the radius of the proposed nuclear ring \citep{wa1995}.

We recognise that the derivation of the kinematic centre, using our large field of view, implies relatively large errors, of 8.6\arcsec\ and 9.2\arcsec\ in the $x$ and $y$ directions, respectively. Allowing for the kinematic centre to vary, the prominence of the circumnuclear rotation velocity rise changes slightly. We tested the rotation curve drop at around 15-20\arcsec\ by displacing the kinematic centre by up to 5 pixels (corresponding to 8\arcsec or 376 pc), after which we derived the rotation curve. For 10 different tests, we found that the location of the drop stays in the radius range 15-20\arcsec, and that the amplitude of the drop-and-rise changes by less than 30 \kms\ (compare with Fig.~\ref{fig:VrotIncl} where the rotation curve drops by about 90 \kms). Thus, changing the centre to within the errors, and re-calculating the rotation curve using the tilted-ring method yields consistent results which confirm that the innermost velocity drop of the rotation curve is genuine.

Centrally rotating structures have been observed in galaxies, and are believed to form as a result of mass concentration caused by a large-scale perturbation or infalling satellites \citep[e.g.,][]{wr1972,Goad1976,bc1977,sn1993,ss1996}. In the case of NGC~628, the clear isolation of this galaxy \citep[e.g.,][]{kb1992} complicates any external formation mechanism for the observed central mass concentration. The large-scale oval structure on the other hand, could transfer matter from the outer zones to regions nearer the centre of the galaxy. Alternatively, the inner disc-like component may be an observational effect caused by a change of inclination between the inner and outer component, i.e., an inner warp \citep[e.g.,][]{ss2006}. Thus the centre could be at higher inclinations, displaying higher deprojected line-of-sight velocities than if it was at the same inclination as the outer parts. As demonstrated in Fig.~\ref{fig:VrotIncl}, the first velocity points of the rotation curve could be at a value lower than (or equal to) the $R$=20\arcsec\ velocity if the inner 20\arcsec, corresponding to 900 pc, would be at an inclination $\simeq 15$ degrees larger than the outer disc. \citet{wa1995} suggested that the nuclear molecular ring might be in a different plane than the outer disc. The nuclear CO deficiency detected by these authors could be only a relative deficiency, since only 45\% of the single-dish flux was recovered in their interferometric observations. In the higher spatial resolution observations by \citet{gandaetal2006}, the nuclear ring appears much more circular, arguing in favour of the nuclear ring residing in the same plane as the outer disc. Moreover, the ellipticity profile of the high resolution $HST$ optical and near-infrared images argue against the presence of a nuclear warp in this galaxy.

The evidence for the presence of a central structure in NGC~628 is not restricted to our \Ha\ velocity field. The position-velocity diagram presented in \citet{daigleetal2006a} clearly shows the presence of a rapidly rotating component in the centre of this galaxy. In addition, our isophotal analysis of archival $I$-band and $H$-band $HST$ data confirm the presence of a nuclear component within the central 20\arcsec\ with a surface brightness profile distinctly deviating from that of the outer parts. Moreover, the $B-V$ colour profile as well as the photometric study by \citet{npr1992} display the presence of an inner component (see our Fig.~\ref{fig:VrotIncl} as well as Fig.~7 of Natali et al. 1992). The latter authors interpreted the blue colour inside the $\sim20$\arcsec\ radius, as an indirect indication of infall of gas to the circumnuclear regions.  Investigating the archival images, we also confirm the presence of the nuclear spiral arms found by \citet{martinietal2003}. Moreover, as we describe in section~\ref{sec:inflow}, our kinematic data does indicate the presence of the structures necessary to induce infall of gas to the circumnuclear regions.

We further investigated the \sauron\ gas and stellar kinematics and distribution maps by \citet{gandaetal2006}, and found confirmation for the $15-20\arcsec$ nuclear ring previously detected by, e.g., \citet{wa1995} in their \Hb\ as well as \OIII\ emission line distribution. The \sauron\ gas velocity field displays large line-of-sight velocities inside 15-20\arcsec\ radius as well as large \Hb\ gas $\sigma$ in the inner 15\arcsec. The stellar component rotates in the same direction as the gas, and its velocity anti-correlates with the third harmonic term inside this radius, indicating the signature of disc-like kinematics \citep{Gerhard1993,vdmf1993}. The stellar $\sigma$ displays a significant decrease of at least 30 \kms\ in the region inside 15\arcsec, qualifying NGC~628 as a $\sigma$-drop candidate \citep[e.g.,][]{emsellemetal2001b}. 

A decrease of the stellar $\sigma$ values in the centres of galaxies requires a distinct dynamically cold and compact isothermal nucleus in addition to the large-scale stellar disc, both with the same mass-to-light ratio \citep[e.g.,][]{Bottema1989}. This picture requires that substantial amounts of gas are directed towards the central regions, the gas cools and forms stars. The newly-born stars are thus concentrated toward the centre, and since they are formed from the low-dispersion gas component, their $\sigma$ is lower than that of the old stellar component \citep{allardetal2005,barbosaetal2006,jfbetal2006,fathietal2006}. This is quite plausible under the circumstances that gas accumulates in a nuclear disc or a bar, where dissipative cooling can be quite effective in producing the condensation required to form stars. This can be a long-lived phenomenon \citep{wc2006}. 

The kinematic as well as photometric signatures observed in the innermost kpc of NGC~628 suggest the presence of a dynamically cold disc-like structure which could be built by inflow from the outer parts towards the centre.

\subsection{Inflow, Circumnuclear Star Formation, and Pseudo-bulge Formation}
\label{sec:inflow}
According to theory, the non-circular motions exerted by a bar or spiral arms funnel material towards the centre of their host galaxy. The gas accumulates in the inner Lindblad resonance region, where circumnuclear star formation can ensue, leading to a low luminosity starburst of up to about  10 M$_\odot$/yr \citep[e.g.,][ and references therein]{knapenetal1995,Athanassoula2005b}. 

\begin{figure}
\includegraphics[width=.49\textwidth]{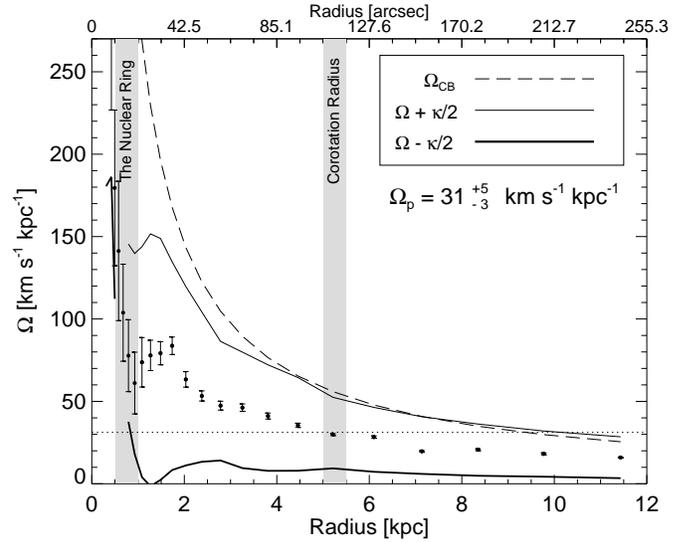}
\caption{The angular frequency, $\Omega$, derived from our observed rotation curve (points with error bars), and that for a flat rotation curve of 33 \kms\ (dashed line) corresponding to the curve drawn by \citet{cb1990}. Also drawn are the $\Omega+\kappa/2$ and $\Omega-\kappa/2$, the latter showing that the location of the nuclear ring is consistent with the location of the inner Lindblad resonance. The horizontal dotted line indicates the value of the pattern speed.}
\label{fig:omega}
\end{figure}

The corotation radius of the spiral arms in NGC~628 was found by \citet{cb1990} as the radius where the star formation efficiency is at its minimum. Using a projected flat rotation curve of 33 \kms, these authors derived a spiral pattern speed of $56$ \kmskpc. We investigate this value by deriving the angular frequency from our tilted-ring rotation curve (see Fig.~\ref{fig:omega}). Considerable discrepancy between our observed curve and a flat one is found. For the corotation radius of $110\arcsec \simeq 5.2$ kpc, our angular frequency curve corresponds to the density wave pattern speed $\Omega_p = 31$ \kmskpc, i.e., almost a factor 2 lower than the value from \citet{cb1990}. An error of 15\% in the presumed location of the corotation radius translates to pattern speed errors of $+5$ and $-3$ \kmskpc. This is clearly due to our improved derivation of the rotation curve, different from that assumed by \citet{cb1990}.

We follow the observed signatures of an $m=2$ perturbation and, using epicyclic approximation, derive the $\Omega\pm\kappa/2$ curves from our angular velocity curve. Given that this curve is rather bumpy, causing artifacts in calculating the numerical derivatives, we derive and present the smoothed $\Omega\pm\kappa/2$ curves. Figure~\ref{fig:omega} shows that our derived pattern speed together with the $\Omega-\kappa/2$ implies the location of the inner Lindblad resonance radius at the position of the nuclear ring as observed by \citet{wa1995} and by \citet{gandaetal2006}. However, the nuclear ring is not fully outlined in \Ha, although two large \HII\ region complexes can be seen at its radius, the gas velocity dispersion rises slightly inside it, and there is a clear deficiency of \HII\ regions just outside it (see Fig.~\ref{fig:sigmaprofile}).

Our \FP\ observations of NGC~628 and the current analysis complement the already existing ingredients which allow us to build an evolutionary scenario for this galaxy. NGC~628 hosts a large-scale oval structure (or weak bar), one regular and a number of disturbed spiral arms,  a nuclear ring at the inner Lindblad resonance radius, an inner, rapidly rotating, disc-like structure, and a nuclear/secondary bar \citep{laineetal2002}, all signs of secular evolution \citep{kk2004}. This galaxy has, most likely, been exposed to some interaction more than 1 Gyr ago, which could cause the disturbed morphology in the south-eastern part of the galaxy. However, the time scale for this interaction is too short for its corresponding perturbation effects to reach the centre. The central parts of this galaxy display signatures of being built up by internal gravitational effects of the prominent $m=2$ perturbation, as a stellar $\sigma$ drop is clearly observed. The young circumnuclear stellar populations observed by \citet{cornettetal1994}, together with the dynamically cold centre invite the interpretation that we are observing the first phases of secular pseudo-bulge formation in NGC~628.

Pseudo-bulges are components intermediate between spheroids and discs. They have shallower light profiles than spheroids, but steeper than discs. Galaxies with pseudo-bulges more frequently host starbursts and circumnuclear star formation \citep{kk2004}. Nuclear star formation in the form of nuclear rings is also preferentially found in barred galaxies \citep[e.g., ][]{knapenetal2006}. \citet{jogeeetal2005} have found that barred galaxies have large enough circumnuclear molecular surface densities to trigger their observed circumnuclear starbursts. Molecular gas surface densities reach up to 20\% of dynamical mass in the centre, eventually leading to the formation of a nuclear ring \citep{pm2006}. Such concentration of gas and star formation, which are more common in late-type spiral galaxies, could lead to the formation of a pseudo-bulge \citep[e.g., ][]{fathipeletier2003}. The structural setup of NGC~628 fits well within this general picture, though this galaxy displays no complete circumnuclear star formation. Our photometric study of the archival data for NGC~628 suggests the presence of a central structure, which together with our kinematic analysis confirm that this central structure is disc-like and could be built up by inflow from large galactocentric distances. Similar cases, such as NGC~4593 and NGC~7690  \citep{kormendyetal2006}, suggest that NGC~628 could be in the process of forming a pseudo-bulge.

\subsection{Velocity Dispersion for the Phases of the Disc Gas}
\label{sec:difflayers}
Different gaseous components of the disc are expected to reach equilibrium. Different scenarios have been suggested for the formation and evolution of different disc layers \citep[][ and references therein]{Kroupa2005}. The forces that govern the small scales are different from those that govern the large scale evolutionary processes. On the smaller scales, dissipation is expected to lower the gas $\sigma$, in order to reach dynamical equilibrium. The discrepant nature of the different gaseous components result in different $\sigma$ levels, and thereby, different thicknesses. Observations have shown that the $\sigma$ of the neutral molecular gas is $\simeq 7$ \kms, $\sigma$ of the warm neutral gas is $\simeq 14$ \kms, $\sigma \leq 30$ \kms\ for warm ionized \Ha-emitting gas, and $\sigma \geq 30$ \kms\ for hot ionized gas \citep{reynolds1985,kf1985}. Hydrodynamic simulations by \citet{dab2005} have shown that these $\sigma$ values remain constant for times as long as 0.5 Gyr, suggesting that thermal and turbulent pressures add to the total pressure within the disc with constant rates. On large scales, gravitational instabilities form an essential heating mechanism, giving rise to vertical heating.

Our analysis of the derived $\sigma$ field of NGC~628 shines some new light on this subject. In Fig.~\ref{fig:sigmaprofile}, we find that $\sigma$ reaches a peak value at around 110\arcsec, which corresponds to the corotation radius suggested by \citet{cb1990}, followed by a slight decrease outside this radius. Taking into account that the variations are within the errors, we find a near constancy of this parameter with respect to galactocentric radius. We find  that the average overall $\sigma$ does not vary when removing pixels belonging to individual \HII\ regions, clearly caused by the smearing effect of averaging along galactocentric rings. Nevertheless, our result confirms that the ionized gas $\sigma$ is higher than its molecular CO and neutral counterparts. We interpret this result as the ionized gaseous component being thicker than the CO molecular and neutral gas components. The puzzling part of this result is the fact that removing the individual \HII\ regions does not change the average $\sigma$ of the \Ha-emitting gas. Could this be due to the highly ionizing effect of individual \HII\ regions on their surrounding interstellar medium?

A substantial amount of star formation is expected in \HII\ regions, the global structure of which remains almost unchanged after a star-forming episode, implying the possibility to study the effect of feedback from stellar winds. Stellar winds not only increase the diffuse interstellar gas $\sigma$, but also contribute to pushing it off the plane of the disc \citep[e.g.,][]{dettmar2004}. One prominent observational example of this effect is the Milky Way, where the scale height of the diffuse interstellar gas is as much as 10 times that of the stellar disc \citep{reynolds1989}, or in NGC~3938, where the ionised gas $\sigma$ has been found to be significantly higher than that of the CO gas \citep{jimenez-vicenteetal1999}. If these three galaxies are typical, they suggest that this effect could be common in spiral galaxies. However more detailed investigation is necessary, and in order to study the feedback effect of the \HII\ regions on the diffuse interstellar medium in NGC~628, one needs to analyse the total emission-line profiles within individual \HII\ regions \citep[see, e.g.,][]{relanoetal2005}. 

It is crucial to quantify the vertical heating efficiency in NGC~628, along with the different emission-line ratios, which enables constraining the thickness of the different gaseous layers. Furthermore, it is necessary to derive gradients in the ionization parameters along the sequence of transition between different states of the gas, and to investigate the underlying physical changes in the ionized gas. Studies of this nature will allow quantifying the energy input from shock-excitation compared with that from photoionization. The currently available data set does not allow for analysis of this sort.

\section{Conclusions}
\label{sec:conclusion}
We have analysed \FP\ interferometric observations of the \Ha\ emitting ionized gas in the late-type spiral galaxy NGC~628 in order to study its internal kinematics over the full face of the galaxy with good angular resolution. Using a harmonic decomposition method we have detected the kinematic signature of an $m=2$ perturbation and this, together with the form of the arms, and in broad agreement with previous kinematic studies at lower resolution made with neutral gas, confirms that the galaxy is evolving with a minimum of external disturbance. We have been able to derive the pattern speed, $\Omega_p$, of the spiral structure using the assumption that at corotation in the disc there is a minimum in the star formation efficiency, and find a value of $31^{+5}_{-3}$ \kmskpc. This pattern speed value is also consistent with the location of the inner Lindblad resonance radius just below one kiloparsec, where a nuclear ring is observed, favouring the inflowing gas scenario according to theory.

Using a continuum subtracted narrow band image in \Ha\ we have identified 376 \HII\ regions with luminosities higher than  36.9 ranging up to 39.3 dex, and with diameters up to 316 pc (see ``Online Material'' at the end). The lower limit here is observational, conditioned by the angular resolution of the image and the distance of the galaxy. As star formation is very widely distributed within the disc we are not able to distinguish in the present exercise, at least for that part of the disc within the radial range of the spiral arms, between true diffuse gas (as represented in the Galaxy by the Reynolds layer) and a face-on ``carpet'' of unresolved \HII\ regions. One clue here is the fact that the mean velocity dispersion for the ionized gas, 16.7 \kms, (and almost constant over the radial range sampled, out to 12 kpc), is hardly affected when the individually detected \HII\ regions are masked off, with a mean value in this case only 0.3 \kms\ smaller. This strongly suggests that in this face-on object with widely distributed star formation in the disc plane, the emission in \Ha\ from the \HII\ regions dominates any emission from the diffuse component.

A key feature of the velocity field of NGC~628 is its regularity. The global effects of any asymmetries such as the oval distortion detected are small, and in any case not very easy to detect in a face-on system. However we clearly detect kinematically an inner disc-like component in the inner kpc around the nucleus, a finding which is supported by corresponding variations in the $B-V$ profile in this zone reported previously. There is at present no excess of star formation within this disc. Our kinematic analysis suggest that the evolution of structure in NGC~628 has been driven by secular evolution of the disc in this galaxy.

\begin{acknowledgements}
We thank Joanna L. Goodger for taking the \Ha\ imaging observations of NGC~628 and for  helping on the \HII\ region catalogue production. The JKT is operated on the island of La Palma by the Isaac Newton Group in the Spanish observatorio del Roque de los Muchachos of the Instituto de Astrof\'\i sica de Andaluc\'\i a. K. Fathi acknowledges support from Wenner-Gren Foundations and the Royal Swedish Academy of Sciences. A. Zurita acknowledges support from the Consejer\'\i a de Eduaci\'on y Ciencia de la Junta de Andaluc\'\i a. Finally, we thank the anonymous referee for insightful comments which improved this manuscript.
\end{acknowledgements}

\bibliographystyle{bibtex/aa}
\bibliography{LR_FathietalN628}

\newpage

\Online

\longtab{1}{
\begin{longtable}{rcccc}
\caption{Complete list of the photometrically calibrated 376 individual \HII\ regions as described in section~\ref{sec:HII}. The luminosities are given in units of erg\,s$^{-1}$, and the coordinates in Equ J2000. Note that here we only list a few of the brightest \HII\ regions. The entire catalogue is made available in electronic format at the CDS via anonymous ftp to {\tt cdsarc.u-strasbg.fr} or via {\tt http://cdsweb.u-strasbg.fr/}.}\\
\hline\hline
Number& R.A. & Dec. & $\log (L)$ & radius (arcsec) \\
\hline
\endfirsthead
\caption{continued.}\\
\hline\hline
Number& R.A. & Dec. & $\log (L)$ & radius (arcsec) \\
\hline
\endhead
\hline
\endfoot
  1   \ \ \ & 01$^h$ 36$^m$ 45.27$^s$	& +15$^d$ 47$^m$ 47.79$^s$  & 39.2577 & 4.456 \\ 
  2   \ \ \ & 01$^h$ 36$^m$ 38.68$^s$	& +15$^d$ 44$^m$ 22.33$^s$  & 39.2266 & 4.204 \\ 
  3   \ \ \ & 01$^h$ 36$^m$ 38.89$^s$	& +15$^d$ 44$^m$ 22.98$^s$  & 39.1637 & 4.091 \\ 
  4   \ \ \ & 01$^h$ 36$^m$ 44.68$^s$	& +15$^d$ 44$^m$ 59.91$^s$  & 39.1053 & 4.378 \\ 
  5   \ \ \ & 01$^h$ 36$^m$ 37.64$^s$	& +15$^d$ 45$^m$ 08.28$^s$  & 39.0848 & 3.667 \\ 
  6   \ \ \ & 01$^h$ 36$^m$ 35.77$^s$	& +15$^d$ 50$^m$ 07.07$^s$  & 39.0790 & 3.742 \\ 
  7   \ \ \ & 01$^h$ 36$^m$ 36.83$^s$	& +15$^d$ 48$^m$ 03.45$^s$  & 38.8143 & 3.214 \\ 
  8   \ \ \ & 01$^h$ 36$^m$ 40.91$^s$	& +15$^d$ 48$^m$ 47.68$^s$  & 38.7797 & 3.427 \\ 
  9   \ \ \ & 01$^h$ 36$^m$ 29.10$^s$	& +15$^d$ 48$^m$ 21.40$^s$  & 38.7429 & 2.391 \\ 
  10  \ \ \ & 01$^h$ 36$^m$ 53.07$^s$	& +15$^d$ 48$^m$ 04.51$^s$  & 38.6818 & 2.890 \\ 
  11  \ \ \ & 01$^h$ 36$^m$ 32.85$^s$	& +15$^d$ 48$^m$ 09.78$^s$  & 38.6641 & 3.148 \\ 
  12  \ \ \ & 01$^h$ 36$^m$ 39.31$^s$	& +15$^d$ 48$^m$ 59.27$^s$  & 38.6097 & 2.914 \\ 
  13  \ \ \ & 01$^h$ 36$^m$ 42.32$^s$	& +15$^d$ 48$^m$ 16.92$^s$  & 38.6093 & 2.497 \\ 
  14  \ \ \ & 01$^h$ 36$^m$ 36.64$^s$	& +15$^d$ 46$^m$ 31.91$^s$  & 38.5934 & 2.566 \\ 
  15  \ \ \ & 01$^h$ 36$^m$ 43.01$^s$	& +15$^d$ 48$^m$ 18.56$^s$  & 38.5883 & 2.973 \\ 
$\vdots$\ \ \ \ \ & $\vdots$                  & $\vdots$                  & $\vdots$& $\vdots$ \\
  376 \ \ \ & 01$^h$ 36$^m$ 58.89$^s$	& +15$^d$ 46$^m$ 22.93$^s$  & 36.8934 & 0.744 \\ 
\end{longtable}
}

\end{document}